\documentstyle[prl,aps,psfig]{revtex}
\bibstyle{unsrt}

\tighten
\begin{document}
\topmargin=0.1cm
\draft

\preprint{MIT-CTP-3220,SUSX-TH-02-001}
\twocolumn[\hsize\textwidth\columnwidth\hsize\csname 
@twocolumnfalse\endcsname

\title{The role of point-like topological excitations at criticality:
from vortices to global monopoles}
\author{Nuno D. Antunes$^{1}$,  Lu\'{\i}s M. A.  Bettencourt$^2$,  
and Martin Kunz$^3$}
\address{$^1$Centre for Theoretical Physics, University of Sussex,
Falmer, Brighton BN1 9QJ, U.K.}
\address{$^2$Center for Theoretical Physics, Massachusetts Institute of 
Technology, Bldg. 6-308, Cambridge 02139}
\address{$^3$Department of Astrophysics, Oxford University, Keble
Road, Oxford OX1 3RH, U.K.} 

\date{\today}
\maketitle

\begin{abstract}
We determine the detailed thermodynamic behavior of vortices in 
the $O(2)$ scalar  model in $2D$ and of global monopoles in the $O(3)$ model 
in $3D$. We construct new numerical techniques, based on cluster decomposition 
algorithms,  to analyze the point defect configurations.  
We find that these criteria produce results for the
Kosterlitz-Thouless temperature in agreement with a topological
transition between a polarizable insulator and a conductor, 
at which free topological charges appear in the system. 
For global monopoles we find no pair unbinding transition. 
Instead a transition to a dense state where 
pairs are no longer distinguishable occurs at $T<T_c$, without leading to 
long range disorder. 
We produce both extensive numerical evidence of this behavior as well as a 
semi-analytic treatment of the partition function for defects. General 
expectations for $N=D>3$ are drawn, based on the observed behavior.    
\end{abstract}

\pacs{PACS Numbers : 64.60.Fr, 11.27.+d, 75.10.Hk, 98.80.Cq \hfill
     MIT-CTP-3220, SUSX-TH-02-002}

\vskip2pc]
     
\section{Introduction}
\label{secI}
     
The many roles played by non-perturbative topological excitations 
in the dynamics and thermodynamics of statistical models and field 
theories is one of the most fascinating and largely unresolved 
issues of many-body systems.

In the simplest Abelian gauge field theories vortices are intimately 
connected with the existence of a superconducting state in type II materials 
\cite{TypeII}. Excitations carrying topological numbers (instantons, 
monopoles, vortices) are also thought to be the best candidates for an 
explanation of confinement in non-Abelian gauge theories \cite{QCD}. 
In the context of statistical models with global symmetries it has long been 
understood that topological excitations can lead  
to the onset of phase disorder at high temperatures \cite{Nelson}. 
Their presence in configurations in 1 spatial
dimensions ($1D$) in models with local interactions down to $T=0^+$ 
prohibits, in fact, the establishment of long range order at any finite 
temperature \cite{Landau,MW}. 
In dissipative dynamical systems the long range disorder and temporal scaling 
present in the long time limit of phase ordering kinetics can also be 
understood in terms of topological excitations \cite{Bray,Mazenko}.   

Several canonical examples illustrate the role of topological excitations 
in bringing about phase transitions \cite{Nelson,Kleinert,CL}.
Among them the best known is the Kosterlitz-Thouless (KT) transition. 
At low temperatures the $O(2)$ model in $2D$ exhibits algebraic 
long range order, which would persist to all temperatures in the absence
of topological excitations. The  advent of disorder at high temperatures 
is due to vortex excitations, which appear as free charges at 
the Kosterlitz-Thouless temperature $T_{\rm KT}$.

Recently, due largely to increases in computational power and improved 
methods, many of these systems have become available to direct quantitative 
study. This is particularly true of  models with global symmetries, 
embodied to a large extent by $O(N)$ symmetric magnets or field theories.  

An interesting  question then is: what happens as we progressively stray away 
from cases where topological excitations are known to dominate critical 
thermodynamic behavior? 
In this article we investigate this question for point defects in $O(N)$ 
models in $D$ dimensions, taking as a starting point the Kosterlitz-Thouless 
transition, i.e. the case $D=N=2$ . The successes of the renormalization 
group at characterizing critical behavior in $D > 2$ suggest that physics 
in the critical region is dominated by perturbative excitations 
(spin waves). In particular as $N$ increases, mean field descriptions 
become suitable. In this limit thermodynamic effects due  to topological 
defects are totally unexpected. What then becomes of the topological 
excitations ? Do they disappear from the spectrum as likely fluctuations, 
or do they still occur but in a manner that does not lead to long range 
disorder ?

The answers to these questions are necessary underpinnings for
a general picture of the behavior of topological excitations  
both in equilibrium and as seeds for the formation of topological 
defects upon cooling. The current understanding of the formation 
and evolution of topological 
defects \cite{Kibble,Zurek} in cosmology and in condensed matter 
requires the presence in the disordered phase of fluctuations, which upon 
cooling can result in long lived topological {\it defects}.
Familiar examples are long vortex strings (cosmic strings) or well separated 
monopole/anti-monopole pairs. 
If these configurations are rare in thermal equilibrium,
above the transition, their abundances will be very small 
and short lived upon cooling. Such behavior may have significant 
phenomenological implications and shed new light on old questions such 
as the monopole problem in cosmology \cite{Monopoles} or the planning of 
defect formation experiments in condensed matter systems.

In this paper we study in detail the similarities and differences 
between the statistical behavior of vortices in the $O(2)$ $2D$ model and 
of global monopoles in the $3D$ $O(3)$ model, in thermal equilibrium. 
For reasons that we make clear in Section~\ref{secII} this step, 
between $N=D=2$ to $N=D=3$ straddles the boundary 
between a case where topological excitations drive an order-disorder 
transition (the former) and a case where topological 
excitations may be expected to become thermodynamically irrelevant,
at least for the long distance physics that is characteristic of criticality.

This problem has been investigated in several instances in the past, 
leading to important insights, but a consistent 
picture of the thermodynamics of global monopoles is yet to emerge. 
The strongest evidence for an important role 
played by  monopoles at criticality in the $O(3)$ model in $3D$, 
comes from the study of modified partition functions, which include monopole 
suppression terms \cite{LD,KM}.  Lau and Dasgupta \cite{LD} showed 
that the introduction of one such term, suppressing {\it all} monopole 
fluctuations, results in the disappearance of critical behavior altogether. 
Later Kamal and Murthy \cite{KM} used a different 
monopole suppression term, which allowed monopole anti-monopole pairs 
with lattice space separation only. They found a new second order transition, 
with exponents different from those of the conventional
$O(3)$ universality class. 
Lau and Dasgupta \cite{LD} also claimed that at the critical temperature 
of the conventional $O(3)$ model the temperature derivative of the
monopole density $d\rho/dT$ 
exhibits a divergence, which they argued would signal monopole/anti-monopole
pair unbinding. This claim was also taken up by Huang, Kolke and Polonyi 
\cite{HKP}, who conjectured that the phase transition in the $O(3)$ model 
would then be driven by monopole/anti-monopole separation, in analogy with 
the vortex unbinding that triggers the Kosterlitz-Thouless transition 
in the $O(2)$ model in $2D$. Later the evidence for a diverging $d \rho/dT$
disappeared with a high precision cluster algorithm study 
by Holme and Janke \cite{HJ}, who showed that  $d \rho/dT$ 
behaves like the specific heat, which does not diverge at $T_c$. 
Moreover, Bitar and Manousakis \cite{BM} searched  for unbound monopoles 
by considering phase correlations along closed loops in space. They 
concluded that no such configurations could ever be found, implying  
that  the unbinding of monopoles plays no role in the critical thermodynamics 
of the $O(3)$ model in $D=3$.

This body of evidence paints a complex picture of the behavior of monopoles 
at criticality in the $O(3)$ magnet in $3D$. It suggests that while monopole 
degrees of freedom are important in bringing about disorder with increasing 
temperature and contribute non-trivially to the physics of the 
$N=3$, $D=3$ universality class, they are not essential for the 
establishment of long range disorder. In particular their behavior may 
not be critical at all at $T_c$. Thus, drawing analogies with the 
Kosterlitz-Thouless transition may give a poor guide to their thermodynamics.

The present  paper is dedicated to elucidating some of these questions, 
through the detailed comparative study of the critical behavior of monopoles 
and  vortices in the $O(3)$ Heisenberg magnet and the  $N=D=2$ $XY$ model, 
respectively.  
This article is organized as follows.
In Section~\ref{secII} we review and extend standard free energetic arguments 
for point defects of $O(N)$ scalar models in $D$ dimensions. These arguments
are both simple and very powerful in determining whether topological 
transitions can occur and in elucidating their nature. 
In Section~\ref{secIII} we characterize the thermodynamic 
behavior of vortices in the Kosterlitz-Thouless transition, by analyzing their
statistical clustering properties. 
We find, in agreement with the Kosterlitz-Thouless paradigm, that 
the transition proceeds by pair unbinding, which can be observed prior 
to a vortex percolation transition. The latter occurs at a slightly higher 
temperature, approximately where the specific heat peaks. 
Armed with this quantitative information of the KT transition and 
the analytical arguments of  Section~\ref{secII}, we analyze, 
in Section~\ref{secIV}, the statistical properties of monopoles 
in the $O(3)$ model in $3D$. 
We find no unbinding transition as expected on energetic grounds alone, 
but still a percolation transition occurs, at a temperature below $T_c$. 
We develop methods to describe the monopole behavior quantitatively and 
argue that the observed percolation transition can occur without leading 
to long range disorder of the order parameter. We thus establish the 
separation between the non-trivial monopole thermodynamic behavior 
and criticality.  
Finally in Section~\ref{secV} we discuss our results in the larger context 
of $O(N)$ scalar theories in $D$ dimensions. We argue that the criteria of 
Section II are sufficient to determine whether topological defects undergo 
an unbinding transition.

\section{Free energy considerations for $O(N)$ point defects 
in $D$-dimensions}
\label{secII}
    
We can gain insight into the importance of topological excitations 
in $O(N)$ models in $D$ dimensions as vehicles of disorder 
by estimating the free energy associated with new pair excitations. 
Later we will specialize to two particular cases, those of vortices in 
$2D$ and of global monopoles in $3D$, whose thermodynamics we investigate 
in detail in Section \ref{secIII}. 
The line of argument used in this section follows the original reasoning  
by Kosterlitz and Thouless \cite{KT} in motivating the topological
transition in the $2D$ $XY$ model, with appropriate generalizations. 

To be definite we consider a general $O(N)$-symmetric $\lambda \phi^4$ 
theory in $D$-spatial dimensions. The Hamiltonian is written as  
 \begin{equation}
 {\cal H}[\phi] = \int d^D{\bf x} \left\{
 \frac{1}{2} |\nabla \phi({\bf x})|^2 +\frac{\lambda}{4} 
\left( \vert \phi({\bf x}) \vert^2 
 -\eta^2\right)^2 \right\},
 \label{hamiltonian}
 \end{equation}
where $\phi({\bf x})$ is a $N$-component real field and 
$\vert \phi \vert^2 = \phi(x)\phi^T(x)$.

The $O(N)$ symmetry of (\ref{hamiltonian}) breaks spontaneously at low 
temperatures to $O(N-1)$, and the field acquires a non-zero expectation value. 
The degenerate set of minima lie on a $S_{N-1}$ sphere. It follows 
that the homotopy group $\pi_{N-1} (S_{N-1}) = Z$, the group of integers, 
which implies that topological solutions with integer charge exist 
in the spectrum of the theory. 
In $D=N$ dimensions these are point defects. The best know cases are the 
kink (or domain wall) in $D=N=1$, the global vortex for $D=N=2$ and the 
global monopole (or hedgehog) in $D=N=3$.  
        
These topological defects are classical static solutions, i.e.  
they are (local) energy minima satisfying $\delta H /\delta \phi = 0$.
In $D$ dimensions point defects are radially symmetric solutions.
The integer topological charge of these configurations implies a singularity 
at their origin ($r=0$), which forces the field amplitude 
$\varphi(r \rightarrow 0) \rightarrow 0$.  For large $r$ it is energetically 
necessary that the field amplitude approaches the minimum of the potential 
$\varphi(r \rightarrow 0) \rightarrow \varphi_0= \eta$.

These boundary conditions do not guarantee that the defect's energy 
is finite. In fact for $D\geq 2$ the energy of topological defects 
still diverges in the infinite volume limit as a consequence of the 
phase gradient terms, which dominate the energy far from the singularity 
at $r=0$. These phase gradients lead to an asymptotic form of the 
energy, for large $l$
\begin{equation}
E \simeq \int_{|{\bf x}|<l} d{\bf x} \, \frac{1}{2} |\nabla \phi|^2
\propto \varphi_0^2 n_q^2 \int^{l}_0 dr \, r^{D-1} \frac{1}{r^2},
\label{En_integral}
\end{equation} 
where $n_q$ is the topological charge of the field configuration.

The diverging energy of a single defect for $D\geq 2$ prohibits it from 
occurring as a fluctuation in thermal equilibrium in the infinite volume 
limit. Instead, defects can occur in defect/anti-defect multipoles 
(usually pairs), which due to mutual screening can then have finite 
energy, a continuous function of their separation. In the case of a pair, the
charge separation introduces a natural cutoff to Eq.~(\ref{En_integral})
which can be used as an estimate for the energy of a pair of size $l$.
 This naive expectation may be changed for $N=D>3$ \cite{Ostlund}, where the minimal energy 
configuration between two topological defects was argued to be one in which 
the far field is rotated to a uniform phase everywhere in space, 
apart from the region between the defects, where energy is concentrated and which 
behaves as a string \cite{Achucarro}. Then the 
interaction potential between two point topological defects in $D=N>3$ 
will be of the same qualitative form as in $D=N=3$, although the associated 
string tension will differ quantitatively (it is expected to  increase 
with $N$).
    
The simplest interesting example of interacting point defects is that
of vortices in $D=2$. The vortex 
anti-vortex dipole has a field 
\begin{eqnarray}
V_{D=2}^{\rm dipole} (r) && \simeq - \varphi_0^2 n_q^2 
\left[ \log(\vert \vec r+\vec l/2 \vert)
-\log(\vert \vec r
-\vec l/2 \vert) \right] \label{dipole_XY}\\ 
&& = -\varphi_0^2 n_q^2 \left[ {\vec r.\vec l \over r^2} 
+ O({l^2\over r^2}) \right]. \nonumber 
\end{eqnarray}
where $\vec l$ is the vector connecting the positive to the negative
charge in the pair.
As a consequence of (\ref{dipole_XY}) a point vortex far away from the dipole 
interacts with it via a potential inversely proportional to distance.
Two well separated pairs then interact with a potential
\begin{eqnarray}
V_{D=2}^{\rm pairs} (r) && \simeq 
\varphi_0^4 n_q^4 \left[ {\vec l_1 . \vec l_2 - 2 (\hat r.\vec l_1) 
(\hat r . \vec l_2) \over r^2} + O({l^3\over r^3}) \right],
\end{eqnarray}
where $\vec l_1$ and $\vec l_2$ are the separation vectors  
within each of the pairs, $\vec r$ is the vector connecting the center 
of the two pairs and  $\hat r = \vec r/r$.
Thus a dilute gas of weakly interacting vortex pairs can exist at low temperatures.
  
Global monopoles have stronger linearly confining potentials. Their dipoles 
therefore behave as
\begin{eqnarray}
V_{D=3}^{\rm dipole}(r)&&  \simeq -\varphi_0^2 n_q^2 \left[ \vert \vec r+\vec l/2 \vert
- \vert \vec r-\vec l/2 \vert  \right] \\
&& = -\varphi_0^2 n_q^2 \left[ {\hat r}.\vec l 
+ O({l^2 \over r}) \right], \nonumber 
\end{eqnarray}
Although finite the potential of this dipole is still substantial.
Two well separated pairs of monopoles then interact via
\begin{eqnarray}
V_{D=3}^{\rm pairs}(r) && 
\simeq \varphi_0^4 n_q^4 \left[ {\vec l_1.\vec l_2 
- ({\hat r}.\vec l_1)({\hat r}.\vec l_2) \over r}   
+ O({l^3 \over r^2}) \right]. 
\end{eqnarray}
Since the interacting potential decreases inversely with separation we see 
that monopole pairs can similarly exist in a dilute, weakly interacting state,
but also that their mutual interactions are stronger than between vortex pairs.
As we will see below this characteristic affects the thermodynamics of 
monopoles relative to vortices considerably. We remark in passing that the 
leading interaction between monopole pairs, apart from polarization inner 
products, behaves in $3D$ as the Coulomb potential. 
The Coulomb gas in $3D$ has a transition, associated with 
the familiar process of ionization, that is a smooth analytical crossover. 
In terms of monopole pairs this transition would occur between a phase 
of free pairs gas and another where monopole pairs become bound to form 
clusters. If it  occurs, we may then expect that this monopole pair 
transition will not lead to critical (i.e. non-analytic) behavior.

To estimate the free energy of a pair of defects we must finally estimate 
its entropy $S=\log(\Omega)$, where $\Omega$ is the number of states of 
the pair. $\Omega$ is proportional to  the surface of the
$D-1$ sphere of radius $l$ i.e. the number of configurations a pair can take 
when rotated around its center of mass.

Then the single pair free energy in arbitrary dimension $D$ is   
\begin{eqnarray}
&F_{D=2} (l) & \simeq E_{\rm c}  + a_D\, \varphi_0^2 \, 
\log(l) - T\, b_D\, \log(l), 
\nonumber   \\
&F_{D>2}(l) & \simeq E_{\rm c}  + a_D\,  \varphi_0^2 \, l^{D-2} - 
 T\, (D-1)\, b_D \,\log(l),\label{Fmonos} 
\end{eqnarray} 
where $E_{\rm c}$ accounts for the total core energy of the two defects in the
pair and $a_D$ and $b_D$ geometrical dimensionless constants dependent
on space dimension.

We emphasize that in these considerations we neglected the effects of other 
defect pairs. Qualitatively these will reduce the free energy of the new 
pair relative to the above estimates, as they will tend to  orient 
themselves in order to (partially) screen the new charges.
Thus (\ref{Fmonos}) should be thought of as an overestimate.

The free energy of the pair gives us a qualitative measure of its 
probability in equilibrium $P(l) \propto e^{- F(l)/T}$. 
We will explore this relationship further in Sec.~\ref{secIII}.
For now we note that for $D=2$ both the energy and entropy terms behave 
logarithmically with $l$ and the overall sign of the free energy for large 
pairs depends on the temperature, as noticed by Kosterlitz and Thouless 
\cite{KT}. In the low-$T$ regime $F$ grows with charge separation leading 
to suppression of large pairs. For high temperatures the entropy term is 
dominant and large pairs have negative free energy. This suggests the 
existence of a high-$T$ phase characterized by the unbinding of 
defect/anti-defect pairs and the production of free charges. 
This is the essential idea behind the Kosterlitz-Thouless mechanism 
for the $O(2)$ $2D$ transition. 
     
For higher dimensions $D>2$ the energy term dominates
at large $l$ for {\em all} temperatures and large pairs always remain 
exponentially suppressed. This simple argument suggests that there
is no unbinding topological transition for $D>2$ in global models and that 
defects remain tightly bound for all $T$. In order to destroy this 
picture it is necessary that the behavior of the bare energy and/or number of 
configurations with $l$ would change due to interactions with other defect
multipoles. Such behavior is not seen in the Kosterlitz-Thouless transition
where the effect of other pairs softens the field modulus $\varphi_0$ 
(the superfluid density or spin wave rigidity), leading to a 
renormalized  value of the transition temperature but not 
to a different kind of transition. In Eq.~(\ref{Fmonos}) we have taken the
pair energy to be determined by a simple cutoff of the single charge total
energy (\ref{En_integral}). Our conclusions would  remain unchanged
if as suggested in \cite{Ostlund} for $D > 2$ the interaction becomes
linear as the field's interacting core collapses into a string connecting the
two charges in the pair. The entropy term would still be dominated by
the interaction energy in the same way as for $D=3$.

Thus we have reached the expectation that no unbinding topological 
transition should ever occur for $N=D>2$. We put this expectation to 
the test in Sec.~\ref{secIII}, where we study comparatively the 
thermodynamic behavior of both vortices in $2D$ and global monopoles in $3D$.  

We devote  the remainder of this section to a few additional remarks 
about the applicability of the free energy considerations of defect 
pairs to more general circumstances. 
An interesting limit is that of systems that remain disordered 
due to topological configurations down to $T=0$. 
In the class of models of (\ref{hamiltonian}) only the case of the 
$N=1$ $\lambda \phi^4$ theory (or the Ising model) in $1D$ has this 
property, due to the presence of kinks (or domain walls). 
In gauge+Higgs field theories the physical properties of topological 
solutions change radically because the phase gradients, 
which dominate the energetics of global defects, become pure gauge 
transformations and carry therefore no energy. This property is a direct 
result of the Higgs mechanism. The phase gradients correspond to 
Nambu-Goldstone modes, each associated with a generator for the remaining 
unbroken symmetries. In the Higgs mechanism these massless modes are 
'eaten' by the gauge field, which in turn acquires a mass for its 
longitudinal projection.
Thus the total energy of gauge defects is concentrated in their cores, 
and falls off exponentially with distance. 
Then gauge topological charges interact via a short range potential, 
in contrast to global defects. This interaction energy can typically 
be neglected or treated effectively, as a change of the core energy, 
in our thermodynamic estimates. 

Either way by repeating the free energy argument with core 
terms only we see that the entropy contribution is dominant for 
all temperatures, 
for large separations in $D \geq 2$. 
This implies that  for large enough separation 
defect/anti-defect pairs are always likely fluctuations, 
and suffer no Boltzmann suppression down to $T=0^+$.
This 'condensation' of free topological excitations can explain striking 
properties of non-Abelian gauge theories \cite{Polyakov}.   
The thermodynamic spectrum of these models should then be characterized 
by the existence of a dilute gas of free defects at low temperatures. 
The numerical verification of this expectation is presently the subject of 
intense research. Hints of this behavior have recently been found numerically, 
see e.g. \cite{Artoo}, for the case of a $SU(2)$ lattice-gauge
theory in $3D$, which possesses 't-Hooft-Polyakov magnetic 
monopoles - the direct analogues of the global monopoles 
in (\ref{hamiltonian}) for $N=D=3$.

\section{Charge Clustering and the Kosterlitz-Thouless Transition}    
\label{secIII}
     
As outlined in Section \ref{secII} the behavior of a system of
charges with logarithmic interactions in $2D$ (the $2D$ Coulomb gas) 
is well known, underpinning the topological transition in the $O(2)$ $2D$ 
model. In this section we develop diagnostics that allow us to measure, 
in numerical studies of the equilibrium partition function, the critical 
properties of topological point charges. Later we will use the 
Kosterlitz-Thouless behavior as the benchmark for a charge unbinding 
transition of $N=D=3$ monopoles.

Rather than using a discretized two dimensional version of the Hamiltonian in
Eq.(\ref{hamiltonian}) we chose to study the $2D$ $XY$ model which belongs
to the same universality class. This choice has the advantage that the $XY$ 
model thermodynamics has  been extensively studied, both analytically
and via large scale numerical simulation.
Consequently its critical properties are well known quantitatively, including
dimensionful quantities such as  the Kosterlitz-Thouless transition
temperature $T_{\rm KT}$ and the temperature at which the specific heat 
peaks $T_{\rm CV}$. 
         
 \begin{figure}
 \centerline{\psfig{file=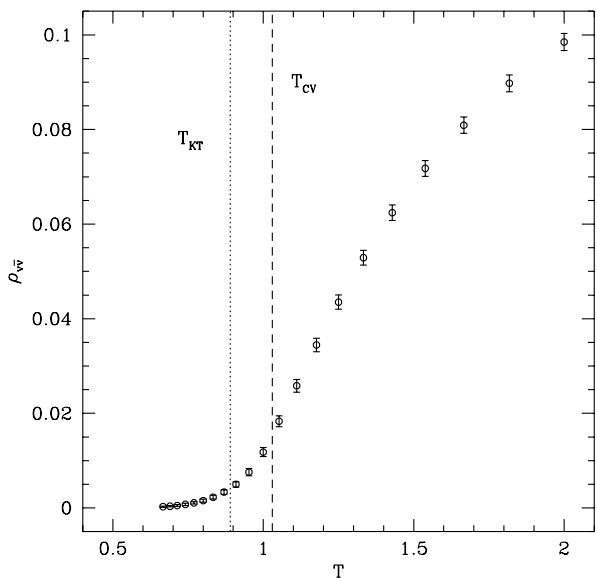,width=3.25in}} 
 \caption{The plaquette density of vortex pairs for the 
 $2D$ $XY$ model. Error bars denote standard deviation over 4000
 independent field realizations. Both the Kosterlitz-Thouless temperature 
 $T_{\rm KT}$ and $T_{\rm CV}$, at which the specific heat peaks, 
 are shown. No signs of critical behavior are apparent in the total 
 vortex density.} 
 \label{rho_XY}
 \end{figure} 
 
The $XY$ model consists of a set of 
two-dimensional unit-length spins with nearest neighbor interaction. 
The Hamiltonian is given by:
\begin{equation}
 {\cal H}[\{{\bf s} \}] = - J \sum_{(i j)} {\bf s}_i \cdot {\bf s}_j,
\end{equation}
where the sum is over all pairs of nearest neighbor sites and we take
$J=1$.

All quantities below were obtained via standard Monte-Carlo generation of large
ensembles on a lattice of size $128^2$. For each temperature we obtained a 
set of $2000$ independent configurations, from which we measured global  
properties of the vortex population. Local quantities not involving
use of time-consuming cluster algorithms were averaged over larger
ensembles. The vortex content of each field 
realization is determined by identifying integer spin windings around the 
lattice plaquettes. The values for the two characteristic temperatures, 
$T_{\rm KT}=0.89$ and $T_{\rm CV}=1.03$  were obtained from the 
literature (see e.g. \cite{Gupta,Olsson}) and are confirmed below. 
In particular we checked that the specific heat peaks at $T_{\rm CV}$
within statistical error.
             
Fig.~\ref{rho_XY} shows  the temperature dependent density of
vortex pairs $\rho_{\rm v {\bar v}}(T)$ defined as the fraction of 
lattice plaquettes occupied by a positive charge.  
Note that although the total vortex density $\rho_{\rm v {\bar v}}(T)$ 
increases with temperature it does not show a clear change 
of behavior at either $T_{\rm KT}$ or  $T_{\rm CV}$. This is not surprising 
since the system does not undergo  a $2^{\rm nd}$-order phase transition, 
and the critical singularities are much weaker in nature. 
In particular the properties of a few large vortex pairs, crucial for 
the onset of phase disorder, are masked in 
$\rho_{\rm v {\bar v}}(T)$ by the existence of many more small pairs. 
In order to see signs of the unbinding we must study the 
properties of the vortex population in greater detail.

To achieve this we must deal with the ambiguity involved in grouping vortices 
in pairs. To overcome this problem in the most general way possible we choose 
to group the vortices in each field realization into 
clusters, defined in terms of an adjustable length parameter $l_{\rm cl}$. 
Vortices or anti-vortices - we do not
distinguish between the two - separated by
less than $l_{\rm cl}$ are collected in the same cluster. Thus, each cluster 
consists of the set of vortices and anti-vortices that lie within
a distance $l_{\rm cl}$ of at least another element of the cluster.

The cluster decomposition is achieved efficiently by applying a 
generalization of the Hoshen-Kopelman algorithm \cite{HK}, developed 
originally for studies of percolation. $l_{\rm cl}$ is successively 
increased, starting from the lattice spacing, the smallest length scale 
in the problem. For each $l_{\rm cl}$  we measure
a set of cluster properties. In particular 
the topological charge properties of clusters are ideal diagnostics 
in the search for signs of a charge unbinding phase transition. 

\begin{figure}
 \centerline{\psfig{file=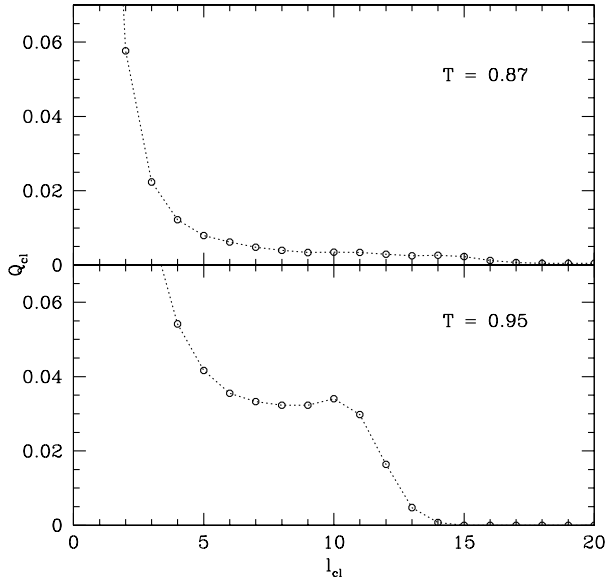,width=3.25in}} 
 \caption{Cluster charge $Q_{\rm cl}$ {\it vs.} clustering length $l_{\rm cl}$ 
 for a temperature slightly below (top plot) and above  (bottom plot)
 the Kosterlitz-Thouless temperature, in 
 the $2D$ $XY$ model. Below the transition the mean charge decreases
 monotonically with $l_{\rm cl}$. The clustering length  $l_{\rm cl}$  
 for which $Q_{\rm cl}=0$ increases as $T \rightarrow T_{\rm KT}^-$. 
 For $T> T_{\rm KT}$, $Q_{\rm cl}$ peaks at an intermediate value 
 of $l_{\rm cl}$ before decaying to zero. Standard deviation error
 bars of order of the results not shown for clarity.} 
 \label{cluster_charge}
 \end{figure}

As a consequence of our choice of  periodic boundary conditions the sum
over the charge of all clusters in the volume is always
zero. To quantify the typical charge of a cluster, 
we define, for each choice of $l_{\rm cl}$, a mean cluster charge $Q_{\rm cl}$.
$Q_{\rm cl}$ results from adding up the absolute value of the charge 
$Q$ of each cluster in a given realization and dividing the result by 
the total number of clusters, i.e.
 \begin{equation}
  Q_{\rm cl} =
     \frac{1}{N_{\rm clusters}}
     \times \sum_{\rm clusters} \vert Q \vert.
 \end{equation}
Fig.~\ref{cluster_charge} shows $Q_{\rm cl}(l_{\rm cl})$  
for two different values of the temperature around $T_{\rm KT}$.
    
The behavior of $Q_{\rm cl}$ at low temperatures can be 
understood in terms of the properties of a dilute distribution of
vortex/anti-vortex pairs. A vortex and an anti-vortex will be in the same
cluster if their separation is smaller than the chosen 
$l_{\rm cl}$. If the distance between 
different pairs is large, all pairs with separation smaller than
$l_{\rm cl}$ will be in neutral clusters. In this case the charged
clusters will consist only of single charges from pairs
with separation larger than $l_{\rm cl}$. With increasing clustering
length, $Q_{\rm cl}$ should then decrease, reaching zero when  $l_{\rm cl}$
becomes of the order of the largest pair in the sample. 
At this length all clusters become neutral. 
We then define $L_{\rm neutral}$ by $Q_{\rm cl} (L_{\rm neutral}) = 0$, 
a measure of the size of the largest pairs at a given temperature. 
     
\begin{figure}
 \centerline{\psfig{file=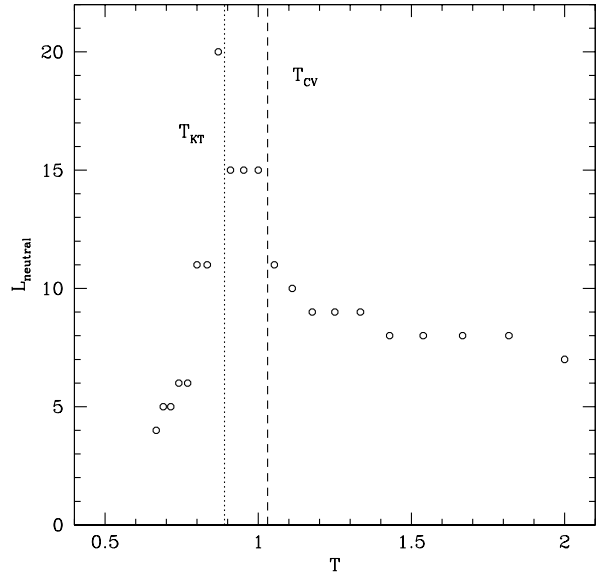,width=3.25in}} 
 \caption{The smallest clustering length $L_{\rm neutral}$, at which all 
 vortex clusters are neutral, {\it vs.} T, in the $2D$ $XY$ model. 
 For low densities $L_{\rm neutral}$ corresponds to the separation 
 of the largest pair in the (finite) ensemble.} 
 \label{length1}
 \end{figure}

The top plot in Fig.~\ref{cluster_charge} illustrates the
typical behavior of $Q_{\rm cl}$ just below $T_{\rm KT}$. 
$Q_{\rm cl} (l_{\rm cl})$ has a long tail for large values of the 
clustering length (up to $l_{\rm cl} \simeq 20$), signalling the presence 
of large pairs. For lower temperatures, $Q_{\rm cl}$ decays 
faster. In Fig.\ref{length1} we plot the variation of $L_{\rm
neutral}$ with $T$. Up to the Kosterlitz-Thouless temperature
$L_{\rm neutral}$ increases as pairs with larger and larger 
separations are produced.
     
According to the standard picture of the KT transition, above $T_{\rm
KT}$ free charges appear in the system. Their presence 
affects $Q_{\rm cl}(l_{\rm cl})$ because a population of free vortices 
changes the charge of otherwise neutral clusters. 
Thus we can no longer expect $Q_{\rm cl}(l_{\rm cl})$ to decay monotonically. 
In fact we observe that above the transition $Q_{\rm cl}$ displays a peak at a
finite value of the clustering length, which we define as $L_{\rm
peak}$. With increasing temperature the value of $L_{\rm peak}$ decreases
and the height of the peak increases. This is the result 
of a higher vortex density and, among them, more free 
charges. The increase in the density of free vortices also reduces the 
mean distance  between them, and moves the peak to lower $l_{\rm cl}$. 
This behavior has the important characteristic that $L_{\rm peak}$ diverges   
as $T_{\rm KT}$ is approached from above. 
Higher free charge densities also lead to a decrease
in $L_{\rm neutral}$ above $T_{\rm KT}$. 

\begin{figure}
 \centerline{\psfig{file=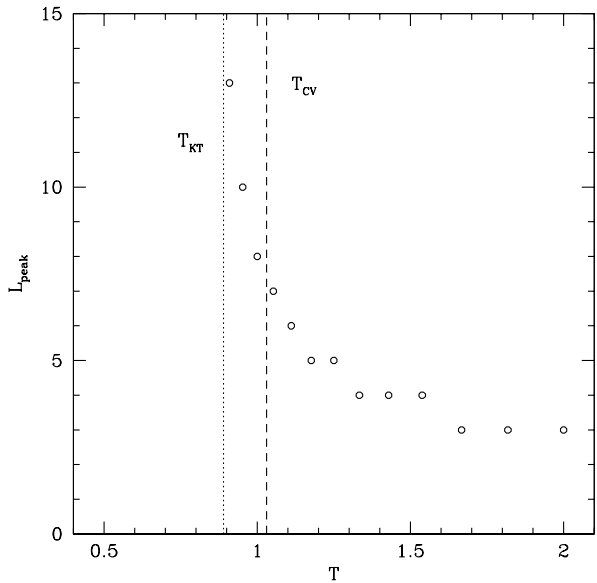,width=3.25in}} 
 \caption{Length for cluster charge peak in the $2D$ $XY$ model. For
 $T > T_{\rm KT}$, the length at which the cluster
 charge function peaks decreases with $T$ due to production of 
 higher densities of bound and unbound pairs. Below $T_{\rm KT}$ the
 mean charge decreases with $l_{\rm cl}$; we define $L_{\rm peak} = \infty$
 in this case.}
 \label{charge_peak}
 \end{figure}

The behavior of both $L_{\rm neutral}$ and $L_{\rm peak}$ is shown in  
Fig.~\ref{length1} and Fig.~\ref{charge_peak}, respectively. 
Both quantities show clear signs of critical behavior at $T_{\rm KT}$, 
although  the behavior of $L_{\rm neutral}$ is plagued by higher statistical 
uncertainty.   
     
Note that pair unbinding is not the only way a set of point charges
may display a change of properties. 
In general as temperature is increased two concurrent effects take place.
The first is that  pairs with larger separation and higher interaction
energy are nucleated. The second is the production of more pairs at 
small separations. Depending on the interplay between  these two trends a 
situation may be reached when  the distance between different pairs 
is of the same order as the separation within each pair. 
In this case the system becomes dense (it percolates) and pairs become  
indistinguishable. 

In order to determine the temperature at which pair percolation occurs 
in the $2D$ $XY$ model  we measured, for each configuration, the value of the 
clustering length at which all clusters become neutral\footnote{
$\langle l_{\rm max}\rangle$ differs from $L_{\rm neutral}$ in the
sense that $\langle l_{\rm max}\rangle$ is a thermal average of the size of 
the largest pair in each sample, whereas $L_{\rm neutral}$ is the size of
the largest pair in all samples in our ensemble. 
In this sense the peak in $\langle l_{\rm max}\rangle$ reflects a maximal 
production of large pairs where the peak in $L_{\rm neutral}$ corresponds 
to the production a single very large (presumably infinite in the infinite 
volume limit) pair.} $\langle l_{\rm max}\rangle$ as well as the 
minimum $l_{\rm cl}$, for which all vortices in the sample fall within
 the same cluster, $L_{\rm perc}$.

\begin{figure}
 \centerline{\psfig{file=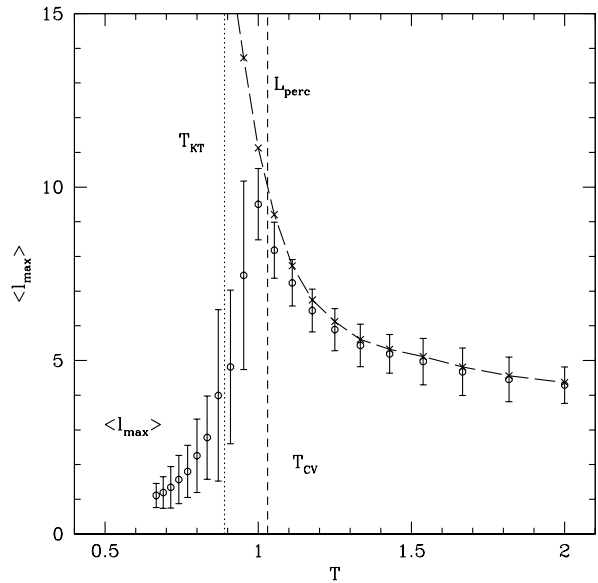,width=3.25in}} 
 \caption{Mean clustering length for neutral clusters $\langle l_{\rm
 max}\rangle$ in the $2D$ $XY$ model, error bars corresponding to
 standard deviation over 2000 field samples. The dashed curve shows
 the clustering length that determines a single cluster - standard
 deviation of the same order as for $\langle l_{\rm max}\rangle$, not shown 
 for clarity. When the two curves meet near $T_{\rm CV}$ the system becomes 
 dense.} 
 \label{length2}
 \end{figure}

For dense vortex systems, where there is no positive 
correlation between vortices and anti-vortices,  
$\langle l_{\rm max}\rangle = L_{\rm perc}$.
This limit must be  reached at high temperature as is indeed 
shown in Fig.~\ref{length2}. 
Moreover we see that for $T < T_{\rm CV}$ the length 
$\langle l_{\rm max}\rangle <  L_{\rm perc}$. This includes the vicinity 
of $T_{\rm KT}$, where pairs remain dilute enough that they can be 
identified. The density threshold where the vortex system becomes 
dense is $T\simeq T_{\rm CV}$, but a precise identification 
would demand careful finite volume scaling. In any case 
we also see that the approach to a dense state occurs seemingly 
continuously, without any clear signs of critical behavior. In this sense 
it may not be possible to associate it with a particular value of $T$.

We have now used cluster decomposition methods applied to the vortex 
population to characterize its critical properties. As the temperature 
is increased we see that free vortices first appear at $T_{\rm KT}$ and 
are maximally produced approximately at $T_{\rm CV}$, where the vortex 
system becomes dense and the concept of a vortex pair ceases to be 
meaningful.

\section{Monopoles in the $3D$ $O(3)$ model}
\label{secIV} 

\subsection{Thermodynamics of the model}

In this section we apply the tools developed in the context
of the Kosterlitz-Thouless transition to a 3D scalar field
theory with $O(3)$ symmetry. Our analysis will be based on a
discretized version of a $\lambda \phi^4$ theory, 
Eq.(\ref{hamiltonian}). We start by establishing its
thermodynamic properties. In particular we will be interested in
determining the value of the critical temperature $T_c$ at which the 
model displays a second order phase transition.
     
In order to generate a Boltzmann distributed ensemble of field
configurations we have evolved a second-order in time Langevin 
field equation (see \cite{ABZ} for more details). For our purposes 
this is equivalent to using a Monte-Carlo or cluster algorithm. The
advantage of the Langevin equation is that it can be easily generalized
for time-dependent systems. In a future publication the equilibrium
states generated in this way will be taken as initial conditions for
real time out-of-equilibrium studies \cite{future}.
     
The procedure is as follows. We evolve the three component
real scalar field in time with the equation of motion
\begin{equation}
\left( \partial_{t}^2 -\nabla^2 \right) \phi_i 
 - m^2 \phi_i \sum_{j=1,3}\phi_j^2 + \lambda \phi_i + 
 \eta \dot{\phi_i} = \Gamma_i, 
\label{e1}
\end{equation}
where $i \in \{1,2,3\}$. We discretize this scheme
using a staggered-leapfrog method 
with time-step $\delta t=0.04$. The random force 
$\Gamma_i(x,t)$ is a Gaussian distributed field
with temperature $T$ as determined by the 
fluctuation-dissipation theorem. $\Gamma_i$ is
characterized by
\begin{equation}
\langle \Gamma_i(x) \rangle = 0, \qquad  
\langle \Gamma_i(x) \Gamma_j(x')\rangle= \frac{2 \eta}{T}  \delta_{ij} 
\delta (x-x').
\label{e2}
\end{equation}
The value of the dissipation coefficient $\eta$ does not influence the
equilibrium results and in our simulations we chose $\eta=1.0$
to ensure rapid convergence. The lattice spacing was set to $\delta
x=0.5$ and the model parameters chosen to be  $m^2=1.0$, $\lambda=1.0$. 
We used a computational domain with $L=100$ points per linear dimension. 
Both local and global observables were measured over at least $200$ independent
field realizations. 
 
As an order parameter we have used the norm of the spatially averaged field 
$\langle \vert \phi_V \vert \rangle$, defined as:
\begin{equation}
\langle \vert \phi_V \vert \rangle = \left\langle \sqrt{\sum_{i=1}^3 \left( 
{1\over{V}}\int_V d{\bf x}~ \phi_{i}({\bf x}) \right)^{2} }\right\rangle,
\label{e3}
\end{equation}
which is analogous to the magnetization in spin models.

 \begin{figure}[ht!]
 \centerline{\psfig{figure=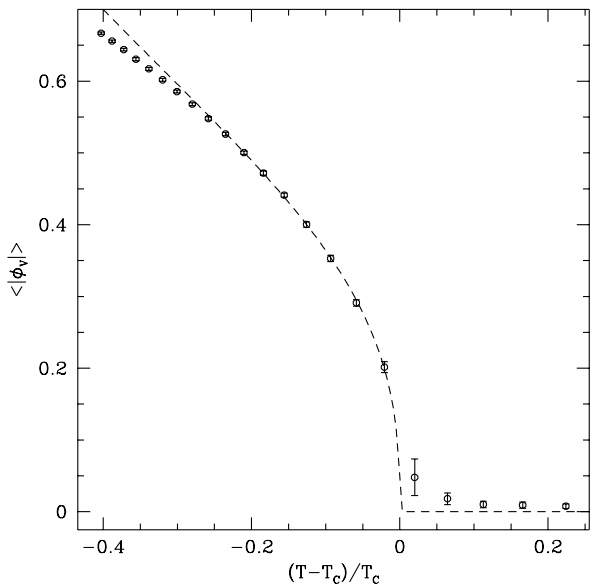,width=3.25in}}
 \caption{Order parameter for the $3D$ $O(3)$ model and corresponding
 power-law fit in the critical region. The temperature has been
 rescaled to $(T-T_c)/T_c$ setting $T_c = 0.41$ as determined from the
 fit. Error bars denote standard deviation over an ensemble of 200
 independent field realizations.}
 \label{ordpar}
 \end{figure} 
          
Fig.\ref{ordpar} shows the temperature dependence of 
$\langle \vert \phi_V \vert \rangle$.
For $T > T_c$, $\langle \vert \phi_V \vert \rangle$
vanishes. Near but below $T_c$, the order parameter
displays universal critical power law behavior
\begin{eqnarray}
\langle \vert \phi_V(T) \vert \rangle = B\;({T_c-T \over T_c})^\beta,
\;\;\;\;\beta>0.
\label{e4}
\end{eqnarray}
which is the analog of the magnetization density in spin models.
Here $\beta$ is the universal critical exponent associated with 
the behavior of the magnetization below $T_c$ and is not to be confused 
to the inverse temperature elsewhere.
By fitting the numerical values for $\langle \vert \phi_V \vert \rangle$
to  Eq.~(\ref{e4}) we are able to measure the critical temperature
obtaining $T_c = 0.41$. This sets a reference point, the most
important scale in the system.
We also compute the critical exponent $\beta=0.36$. 
This is in good agreement with both recent theoretical and
large scale Monte-Carlo estimates for the critical exponent which give
$\beta=0.366(2)$ and $\beta=  0.3685(11)$ respectively 
(see e.g. \cite{crit_phen} and references
therein) and provides a check on the accuracy of our numerical setup.

\subsection{Monopole Statistics}

We are now ready to analyze the properties of the
equilibrium global monopole population. For each field realization 
monopoles and anti-monopoles
are identified by measuring the three dimensional winding number of the
field around each cubic lattice cell. 
The algorithm used is based on a higher dimensional generalization of the 
geodesic rule traditionally used for identifying strings in $O(2)$
theories \cite{VV}. Details of this procedure are given in 
Appendix \ref{winding}.

 \begin{figure}[h!]
 \centerline{\psfig{figure=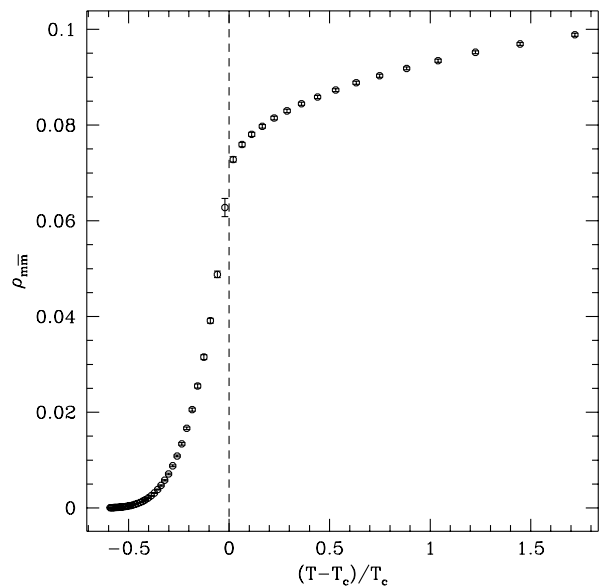,width=3.25in}}
 \caption{The mean plaquette density of monopole pairs 
    $\rho_{\rm m {\bar m}}$ {\it vs.} reduced temperature  
 for the $3D$ $O(3)$ model - error bars as in Fig.~\ref{ordpar}.}
 \label{monopole_density}
 \end{figure} 
     
Fig.~\ref{monopole_density} shows the density of monopole/anti-monopole
pairs (defined as the total positive charge in the computational domain 
divided by the number of lattice sites) {\it versus}  
temperature. The total monopole density $\rho_{\rm m {\bar m}}$ increases 
smoothly with $T$ and its derivative peaks at the
critical point signalling the second-order phase transition.
Above $T_c$ the rate of increase diminishes and the total pair 
density converges slowly to approximately $\rho_{\rm m {\bar m}} 
\rightarrow 0.17$ as $T \rightarrow \infty$ (not shown). 

\begin{figure}[h!]
 \centerline{\psfig{figure=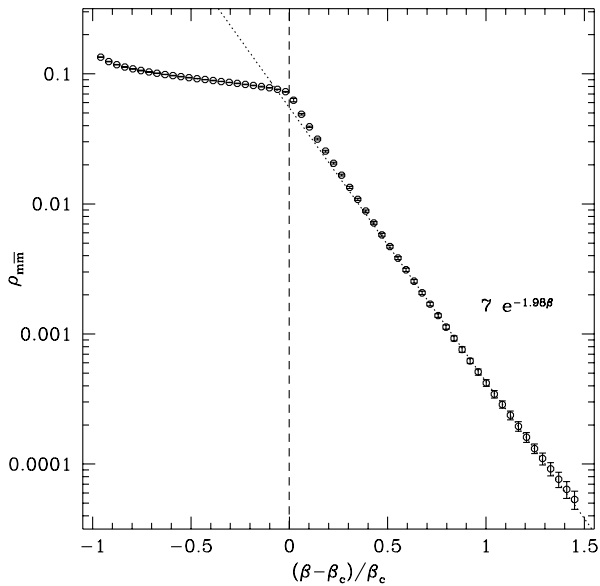,width=3.25in}}
 \caption{Fit for the mean density of monopoles to $A \exp(-
 E_0/T)$, error-bars as above. 
 In the low temperature regime we obtain $A=7.0 \pm 0.6$ and 
 $E_0 = 1.98 \pm 0.02$, fitting the first 23 data points. Below $T_c$
 the density displays near exponential behavior down to $T=0$.} 
 \label{monopole_density_log}
 \end{figure}

Fig.~\ref{monopole_density_log} shows a log-linear
version of Fig.~\ref{monopole_density} illustrating how over nearly the 
whole temperature range below $T_c$, $\rho_{\rm m {\bar m}}$  is well fit 
by an exponential $\rho_{\rm m {\bar m}} = A\, e^{- E_0/T}$. 
Only very near the critical point does the fit fail to follow the density 
curve accurately. This behavior suggests that the increase in
the total monopole density is dominated by the creation of large numbers of 
minimum size pairs, each with typical energy $E_0 \simeq 2.0$.

An understanding of this behavior can be obtained by evaluating the 
partition function for monopole pairs under certain simplifications. 
If we assume pairs are independent i.e. we neglect 
pair/pair interactions and volume exclusion effects,
the partition function for a pair is
\begin{equation}
Z(T) = 1 + \sum_{\rm pairs} e^{- E_p/T},
\label{part_func}
\end{equation}    
where the sum is taken over all single pair internal configurations, i.e. 
it excludes translational modes. We take the pair energy to be of the form
\begin{equation}
E_p = E_c + \sigma l,
\label{E(l)}
\end{equation}
where $l$ is the monopole anti-monopole separation in
units of lattice spacing. In the continuum the simplest way to compute 
$Z(T)$ would be to use the approximate expression for the free energy 
Eq.(\ref{Fmonos}). 
While this should be valid for large values of the pair size, such evaluation 
of the number of states breaks down relative to that on the lattice, 
especially when $l$ becomes of the order of the lattice spacing. 
Since we expect the monopole population to be dominated by small
pairs, the continuum approximation would be a significant source of error. 
To circumvent this problem  we calculate the partition function 
by evaluating numerically the sum in (\ref{part_func}) over all possible 
lattice configurations of a pair with fixed center, for a given choice 
of $E_c$ and $\sigma$. In this way the number of pair configurations on the 
lattice, and therefore the entropy, are calculated exactly. 

 \begin{figure}[h!]
 \centerline{\psfig{figure=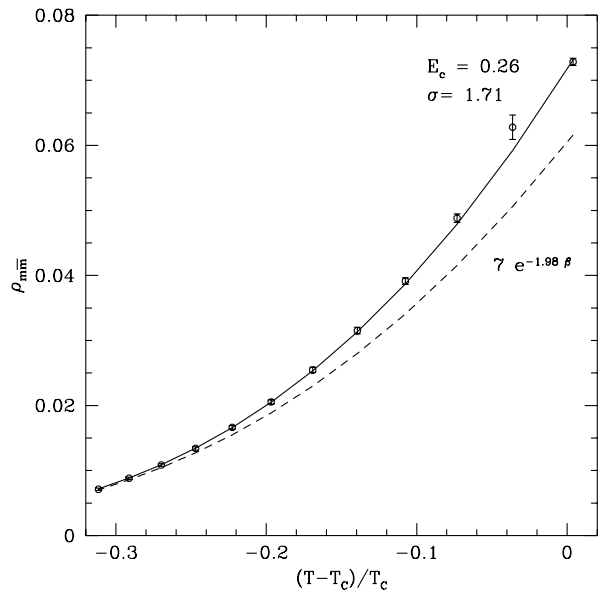,width=3.25in}}
 \caption{Numerical data for the total pair density and
 corresponding fits near $T_c$, error-bars as in Fig.~\ref{monopole_density}.
 The energy varying theoretical curve (solid line) was fitted to the
 18 lowest temperature points (not shown in the plot) up to
 $(T-T_c)/T_c = -0.25$. The simple exponential fit (dashed line) fails
 to follow the numerical
 curve up to $T_c$ indicating relevant production of higher-energy, larger
 separation pairs in the critical region. The varying energy fit on
 the contrary matches the data well up to the critical point.} 
 \label{monopole_fits}
 \end{figure}       

Monopole thermodynamic averages can be easily evaluated from $Z(T)$. 
The total pair density per site is given by:
\begin{equation}
\rho_{\rm m {\bar m}}(T) = \frac{Z(T)-1}{Z(T)}.
\end{equation}
By allowing $E_c$ and $\sigma$ to take arbitrary values, 
we fit the predicted pair density to
the numerical data. A $\chi$ squared fit leads to 
estimated values of the interaction  parameters of $E_c\simeq 0.26$ and
$\sigma \simeq 1.71$. 
          
In Fig.~\ref{monopole_fits} we compare the monopole density 
$\rho_{\rm m {\bar m}}$ obtained in this way to the numerical data and 
to the simple exponential fit discussed 
before. Clearly  taking into account pairs with variable separation 
improves the estimate leading to precise results up to $T_c$. 
In spite of this improvement it remains true
that the monopole thermodynamics is always dominated by very 
small pairs.  Using the fit results for  $E_c$ and $\sigma$ we obtain 
the following energies for the smallest pairs allowed on the lattice: 
$E(1)=1.97$, $E(\sqrt{2})=2.68$ and $E(\sqrt{3})=3.22$. 
Considering the contribution of the first three terms in the partition 
function already leads to a very reasonable approximation to $Z(T)$:
\begin{equation}
Z(T) = 1+6 \, e^{-1.97/T}+12\, e^{-2.68/T}+8\, e^{-3.22/T}.
\end{equation}
This gives $\rho_{\rm m {\bar m}}(T_c) = 0.07$, in good agreement with 
the measured value (the integer pre-factors are the number of different 
lattice configurations for a pair at these separations).
     
The value measured for $\sigma$ is considerably lower than
the one obtained from the single monopole classical estimate in Section
\ref{secII}. Evaluating Eq.~(\ref{En_integral}) exactly we have 
for the single pair energy $E_{p} = E_c + 4\pi m^2/\lambda \times l$, which
leads to $\sigma = 2\pi$ in lattice spacing units 
($\delta x = 0.5$). The difference between this classical value of 
$\sigma$ and its value inferred from fitting the thermodynamic monopole 
density is a consequence of strong medium dressing effects, resulting 
both from the influence of other monopole pairs and from interactions
with the spin wave degrees of freedom. 

In any case the unequivocal exponential behavior of the total density below
the critical point implies that $\sigma \le 2$.
The quality of the fit using $\sigma = 1.7$ and its success in
predicting other features (see below) of the data
suggests that one should not place exaggerated confidence in
the classical single monopole result.

In a previous publication \cite{ABY} two of us predicted 
the value of defect density at criticality for $O(N)$ theories. 
This calculation assumes that fluctuations at $T_c$ are Gaussian,
with their scale invariant connected 2-point function 
characterized by the universal critical exponent $\eta$, 
the anomalous dimension. Using the value of 
$\rho_{\rm m {\bar m}}=0.17$ at infinite temperature as a normalization 
(see \cite{ABY} for details) leads to a predicted 
$\rho_{\rm m {\bar m}}(T_c) \simeq 0.07$ for $O(3)$ in $3D$, 
in good agreement with the present numerical measurements.

\begin{figure}[h!]
 \centerline{\psfig{figure=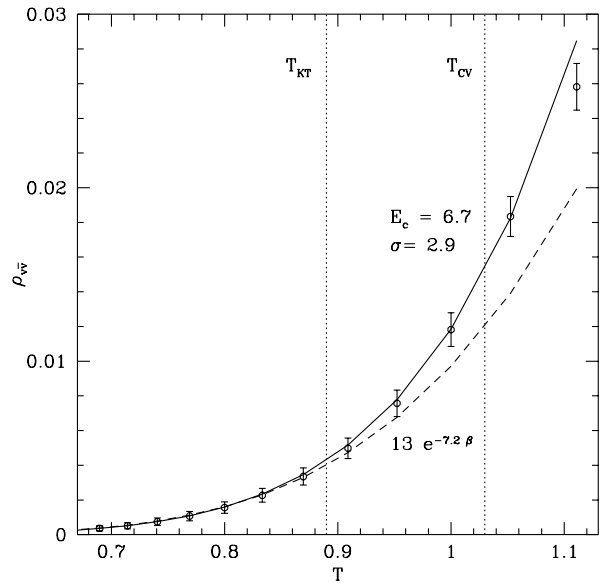,width=3.25in}}
 \caption{Same as Fig.\ref{monopole_fits} for the $2D$ $XY$ model. The
 theoretical curves were obtained by fitting the low-temperature data up
 to $T=0.75$. Whereas the simple exponential (dashed line) fails to follow the
 observed density above $T_{\rm KT}$. The curve obtained from the 
 partition function, which includes pairs with all separations, matches
 the data well up to $T_{\rm CV}.$} 
 \label{vortex_fits}
 \end{figure}

A similar calculation of the temperature dependence of the vortex
pair density can be done for the $2D$ $XY$ model. As
in the monopole case, the low temperature Monte Carlo data is reasonably 
well fitted by an exponential. Assuming a pair energy of the form
$E_p = E_c + \sigma \log(l)$ and calculating the partition function
as before, the prediction can be improved leading to good results up
to $T_{\rm CV}$ (see Fig.~\ref{vortex_fits}). The single exponential
fit for low-$T$ gives $E_0 \simeq 7.2$ which compares well with a
previous result of $7.5$, measured by Gupta and Baillie \cite{Gupta}.
The difference is probably due to our fit being based on
low-$T$ data points only. 
In the same article \cite{Gupta} Gupta and Baillie also obtained 
a different  exponential fit in the temperature region 
between $T_{\rm KT}$ and $T_{\rm CV}$ with a higher value for $E_0$. 
Using the form  (\ref{E(l)}) we are able to fit both temperature regimes 
obtaining the $\chi$ squared results $E_0 \simeq 6.7$ and $\sigma \simeq 2.9$.
     
\begin{figure}[ht!]
  \centerline{\psfig{figure=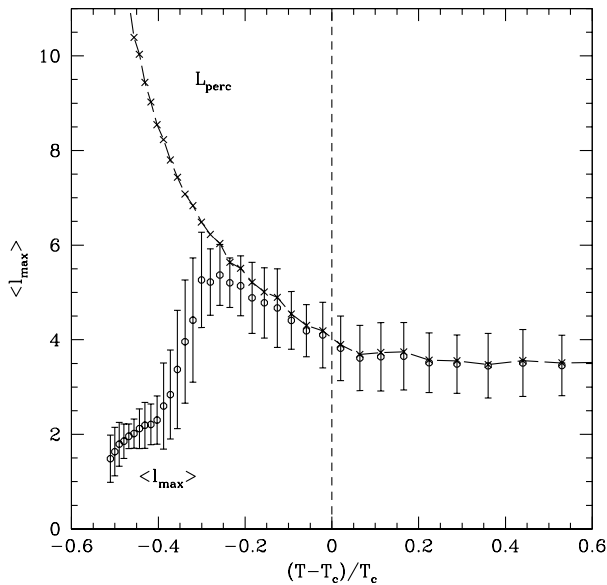,width=3.25in}}
  \caption{Mean clustering length for neutral clusters (error-bars
  correspond to standard deviation over 200 field samples) and percolation
  length (dashed curve - standard deviation error bars of the same order as 
  for $\langle l_{\rm max}\rangle$, not shown for clarity) for the $3D$ 
  $O(3)$ model. 
  The system 
  becomes dense well below the critical temperature when the two
  quantities become comparable.}
  \label{l_monopoles}
  \end{figure}

Following in the footsteps of the $2D$ charge cluster analysis  
we now turn to the properties of monopoles in the $O(3)$ $3D$ 
theory. Fig.~\ref{l_monopoles} shows  $\langle l_{\rm max}\rangle$
and $L_{\rm perc}$ in terms of the reduced temperature 
in the critical region. The monopole ensemble becomes 
dense when the two length scales are comparable, 
$\langle l_{\rm max} \rangle \simeq L_{\rm perc}$ and it is no longer 
possible to identify isolated pairs. 
This happens at $(T-T_c)/T_c \simeq -0.25$, 
the temperature at which $\langle l_{\rm max}\rangle$ peaks, 
well below $T_c$.  This behavior stands in striking contrast to that of  
vortices  in the $2D$ case (see Fig.\ref{length2}), 
where the vortex gas percolated only in the exponentially disordered 
phase at $T=T_{\rm CV}> T_{\rm KT}$. 
    
It is important to realize that this behavior of monopoles is not in 
contradiction with maintaining long range order up to $T_c$. 
The system of monopole anti-monopole pairs can 
become dense without disordering the field over large distances. 
This can be understood by considering a domain with radius much larger 
than the maximal pair size and is essentially the earlier result of 
Bitar and Manousakis \cite{BM}. 
Since the total field winding in the domain's surface is given
by the total charge in its interior, its value will be zero. That is,
finite pairs will not affect the long range behavior of the field 
on scales larger than their size, only unbound charges can lead to the 
break down of long range order. 
 
 The proliferation of small pairs in the three dimensional case is
made possible by the fact that the core energy of the monopole is
small when compared with the interaction potential. This leads to 
production of large densities of small pairs, while large pairs
remain strongly suppressed by the fast growing linear interaction
term. As a consequence the system percolates mostly due to a high
density population of small pairs at a temperature where large
pair configurations have a exponentially negligible contribution to the
thermodynamics.
    
\begin{figure}[ht!]
 \centerline{\psfig{figure=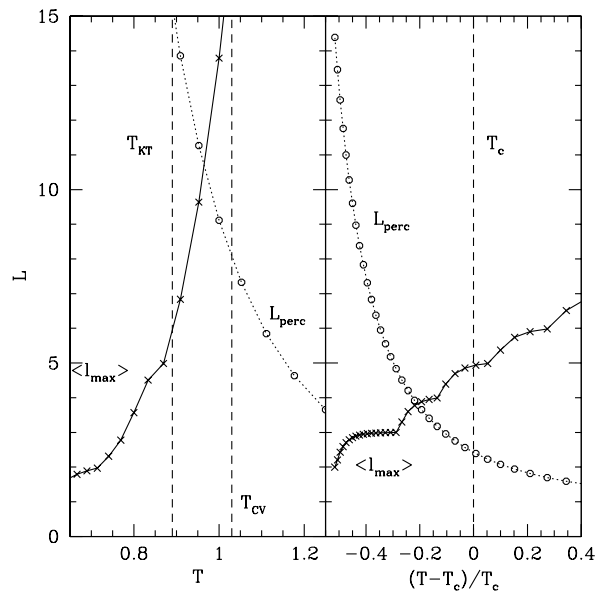,width=3.25in}}
 \caption{Theoretical prediction for the temperature dependent mean maximal
 pair size $\langle l_{\rm max}\rangle$ and pair separation
 $L_{\rm perc}$, for the $2D$ $XY$ model (left) and the $3D$ $O(3)$ model
 (right). At the temperature where the two curves cross, the
 system becomes dense and it is no longer possible to distinguish 
 individual pairs. The step-like appearance of $\langle l_{\rm
 max}\rangle$ in the right plot is due to lattice discretization
 effects on $l$, which are apparent due to the small size of the
 monopole/anti-monopole pairs.}
 \label{2predictions}
 \end{figure}

To investigate this behavior we can use our approximate partition function to 
calculate $N(l)$, the density of pairs of size $l$ as 
 \begin{equation}
 N(l) = Z^{-1}(T)\, n(l)\, e^{-E_p(l)/T},
 \end{equation}
where $n(l)$ is the total number of configurations for a pair of size $l$. 
This expression can be readily evaluated numerically. We then use $N(l)$
to estimate the size of the largest pair in a computational domain
by finding the value for $l$ such that  $N(l) \times L^D \simeq 1.$,
where $L$ is the linear size of the computational volume. 
That is, we demand that in each computational volume there should 
be on average one pair of maximal size. This 
length scale corresponds to $\langle l_{\rm max}\rangle$. 
The value  of $\langle l_{\rm max}\rangle$ estimated in
this way is plotted against temperature for both models in
Fig.~\ref{2predictions}.
     
 In order to calculate the percolation temperature in this
approximation we must estimate $L_{\rm perc}$. This can be done by
assuming that the typical distance between pairs is of
order of $1/\rho^{(1/D)}$, where $\rho$ denotes either 
$\rho_{\rm v {\bar v}}$ or $\rho_{\rm m {\bar m}}$ depending on the 
dimension. This implies that we will have percolation 
when $ \simeq 1/\rho^{(1/D)}$. Fig.\ref{2predictions} shows the 
temperature dependence of $L_{\rm perc}$, estimated as
$1/\rho^{(1/D)}$, using the parameters from the previous density fits. 
The point where the two curves meet defines the percolation
temperature where the change ensemble becomes dense. The values obtained
agree reasonably well with the data. 
For the $3D$ $O(3)$ case we find the percolation temperature to be 
$(T-T_c)/T_c\simeq -0.2$ with pairs of maximal size $4$, compared to 
$(T-T_c)/T_c\simeq -0.25$ with largest length of around $5.5$ from the 
numerical results. For the $XY$ model we obtain $T\simeq 0.96$ 
slightly below $T_{\rm CV}$ but still clearly in the exponentially
disordered phase and $l\simeq 10$, which coincides with the numerical result
for the mean size of the largest pair per box at the percolation temperature.

\section{Conclusions}
\label{secV}

We started this paper by using single defect pair  
free energetic arguments to predict that no pair unbinding transition
should occur for point defects in $O(N)$ models with global 
symmetries, for $N=D>2$.
We then verified this prediction by comparatively studying the defect 
thermodynamics of vortices and global monopoles occurring in typical 
configurations drawn numerically from canonical ensembles of the 
$O(2)$ model in $2D$ and an $O(3)$ field theory in $3D$.

We measured in great detail the behavior of vortices 
across the  Kosterlitz-Thouless transition and found it in 
agreement with the expectations of the theory, which predicts the appearance 
of free vortices at $T_{\rm KT}$. In the $O(3)$ model in $3D$ isolated 
single monopoles never occur. Instead the monopole ensemble transits 
from a dilute pair phase at low temperatures, to a dense monopole gas, through 
the nucleation of large numbers of tightly bound pairs below $T_c$. 
We have seen that this behavior is consistent with qualitative expectations 
based on the single monopole pair energetics, 
which is characterized by a light core and a linearly confining 
interaction potential. Our quantitative
treatment shows, however, that the bare parameters in the monopole 
potential, the monopole core energy $E_c$ and string tension $\sigma$, 
are significantly renormalized  by thermal medium effects. 

The thermodynamic behavior of global monopoles is consistent 
with both a (non-critical) topological charge conductor to insulator 
transition and the absence of long range phase disorder at temperatures 
below $T_c$. 
Global monopoles in $3D$ do not behave like Coulomb charges, but rather more 
like static quarks, since they interact via a linearly confining potential. 
The conductor phase in this case cannot be reached by the nucleation 
of truly free isolated charges, as in the Kosterlitz-Thouless transition. 
It is instead the result from the fact that in a dense monopole gas 
charges can move freely, by hopping from a nearby anti-charge to another. 
Because local charge neutrality persists to all finite temperatures \cite{BM}
the dense monopole gas does not lead to long range disorder of the field 
phases (spins). Long range phase fluctuations must therefore arise from 
different degrees of freedom, the spin waves. 
This is indeed the renormalization group picture 
of the transition, which correctly predicts all universal 
critical behavior. From this perspective we conclude that monopole 
excitations, although contributing to short range disorder in the $O(3)$ 
model in $3D$, are incapable by themselves, because of their enslaving 
energetics, of producing long range disorder at any finite temperature.
This is not to say that their disordering influence can be 
completely neglected. 
Their effects must generate a short distance nontrivial quantitative 
contribution  to critical exponents, thus making the universality class 
in their absence different \cite{KM}. 

In this way we must conclude that there is a non-trivial interplay 
between monopoles and spin waves in the $O(3)$ model and that the detailed 
nature of critical behavior is changed if either are suppressed.

The generalization of these results to arbitrary $N=D>3$ is immediate.
The potential between defects becomes steeper and steeper as a function 
of $l$, and topological excitations, just like monopoles in $3D$, will
never unbind. In this way topological excitations do indeed become 
more and more irrelevant, as $N$ increases, for the physics of large 
length scales that characterizes thermodynamics in the critical 
domain of $O(N)$ models.

We also see from this perspective that, upon cooling the system, 
most topological fluctuations will annihilate with a nearby 
anti-defect, leading to small and quickly disappearing populations
of topological defects. From this perspective global 
topological monopoles formed at a cosmological phase transition
present no real danger of creating a `monopole problem'.
 
We conclude by invoking a complementary view of the phase transition 
in $O(N)$ models. There is clear evidence (and mathematical 
proofs in certain particular cases) that  criticality
in $O(N)$ scalar models is equivalent to the percolation of so-called 
Wolff spin clusters \cite{Clusters}. 
Wolff clusters are built by forming bonds among 
adjacent spins according to a temperature dependent probability. 
Because of this probabilistic assignment  Wolff clusters are subsets 
of the set of clusters formed by associating all adjacent spins with 
the same orientation. Clusters formed by considering spins with the same 
orientation, without this probabilistic restriction, percolate below $T_c$. 
It is the typical size of these latter 'conventional' clusters
that is associated with defect densities, according to the Kibble-Zurek 
\cite{Kibble,Zurek} theory of defect formation. 
Our observation of monopole percolation below $T_c$ is compatible 
with this scenario. The interesting questions of testing this hypothesis 
and of determining the detailed relationship between Wolff clusters and 
topological excitations at criticality will be left for future work.

\acknowledgments
     
 This work is supported in part by the Department of Energy under
 cooperative research agreement \#DF-FC02-94ER40818.   
 N. D. A. was supported by a PPARC Postdoctoral Fellowship.
 M. K. acknowledges support from the Swiss National Science 
 Foundation under contract 83EU-062445. We would like to
 thank R. Durrer for helpful discussions.
 N. D. A. would like to thank G. Volovik and A. Schakel for 
 useful suggestions. 
     
\begin{appendix}
     
\section{Energetics of line defects in $D$-dimensions}
     
  The free energy arguments of Sec.~\ref{secII} can be generalized
  to extended topological defects. In this appendix we consider the 
  case of line defects.

  Let us look first at the case of one-dimensional defects $N=D-1$. 
  For a string-like object the core energy is proportional to its
  length $l$, and the entropy can easily be calculated assuming that
  it behaves like a random walk. The number of different
  configurations for a closed random walk in a cubic $D$-dimensional
  lattice is given by:
  \begin{equation}
  \Omega =  {(2 D)^l}{(4 \pi l)^{-D/2}} \times l^{-1}.
  \end{equation}
  The first two terms count the number of possible closed
  random walk loops with $l$ steps and the final factor takes into
  account the arbitrariness in the choice of initial point in the
  loop. We see that the entropy $S=\log(\Omega)$ grows linearly with $l$. 
  The free energy is then (omitting dimensionless constants):
  \begin{equation}
  F(l) \simeq  \sigma l +  E_I + T (\frac{D}{2}+1) \log(l) - \log(2D) T l,
  \end{equation}   
 where $\sigma$ is the string tension and $E_I(l)$ the interaction energy
 for the loop. In general $E_I$ is a function not only of the string length 
 but also of its detailed shape.
 We start by considering the case when the  interaction energy can be 
 neglected. Then we see immediately that the system will undergo a 
 phase transition above a certain $T$, 
 characterized by the proliferation of  long strings 
 This is because both the entropy and core energy have the same dependence
 on $l$, as was the case in the Kosterlitz-Thouless transition. 

 This set of approximations gives a reasonable description of the $D=2$ 
 Ising model, where there is no long range interaction between strings 
 and probably also of strings in gauge theories in any dimension. 
 In contrast to the case of monopoles, where the pair interaction 
 was responsible for the confining phase at low-$T$, here it is the 
 core energy of the string that keeps it at a finite length.

 The interaction energy is hard to estimate and depends in general 
 on details of the underlying model as well as of particular configurations. 
 We can make a rough estimate by assuming that the string remains a random 
 walk and consequently that the distance between two points on the string 
 is of order of $l^{1/2}$. Then if 
 two distant string segments interact as point-like defects in a
 $O(N-1), D-1$ theory, the loop energy will be given by:
 \begin{eqnarray}
  &E_I(l)& \simeq l\times l^{(D-3)/2},\,\,\,\,D>3\nonumber \\
  &E_I(l)& \simeq   l \log(l),\,\,\,\,D=3   
 \end{eqnarray} 
 Clearly for $D>3$ the interaction term dominates over the entropy
 for all $T$ indicating the absence of a proliferating phase. The
 marginal case, for $D=3$, is harder to judge in this approximation, since
 the correction to the linear interaction is only logarithmic. 
 In reality we have good evidence \cite{ABH} that for $3D$ $O(2)$
 there is a percolation transition in the string network at $T_c$ 
 at which long strings spanning the volume appear in the system.

  For higher dimensional defects the calculation of the entropy
 becomes much harder. For example surfaces can be topologically 
 complicated, exhibiting holes and handles. The general trend 
 seems to suggest that for fixed spatial dimension, the entropy 
 of a typical defect configuration decreases with $N$. Since the 
 interaction energy increases with the number of field components, 
 we expect that for any $D$, and above a certain $N$ the free 
 energy will  always be dominated by the interaction component 
 and proliferation of defects is prohibited at all finite 
 temperatures. For $D=3$, for example, strings proliferate in the 
 high-$T$ phase for $N=2$ but no unbinding of charge pairs occurs for $N=3$.

\section{Defining Winding Charge on the Lattice for Global Theories}
\label{winding}
   
We identify $O(3)$ monopoles in a $3D$ cubic lattice by generalizing 
the well known `geodesic rule' used in most cosmic-string 
lattice based simulations \cite{VV}.
     
\begin{figure}
\centerline{\psfig{file=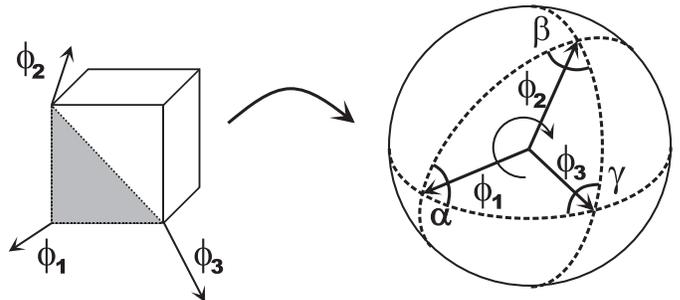,width=3.25in}} 
\caption{The monopole charge in a cubic cell is identified by
projecting the field-vectors at every corner onto a unit sphere.
Each triangle on the square faces cube is thus mapped into a
a spherical triangle (the
one with the smallest surface is chosen). The sum of the surface of
all these triangles, divided by $4\pi$, is then taken as defining the
monopole charge inside the cubic cell.}
\label{fig1}
\end{figure}  

We count the winding of the field-vectors around each unit 
cell in our grid by using a `smallest area' assumption.
To this end, we triangulate the faces
of each lattice cube and then map the $O(3)$ field-vector at
all corners onto the unit sphere (the presence of a monopole 
depends only on the orientation of the field, not on its norm). 
For each triangular element in the cube's surface this defines a 
solid angle on the unit sphere (see fig.~\ref{fig1}). The sign of
the solid angle is taken according to the handedness of the corners. 
Its value $\Theta$ can be calculated thanks to a formula that relates the
area of the spherical triangle, defined by three vectors on a unit
sphere, to the angles between the geodesic sides of the triangle:
\begin{equation}
|\Theta|= \alpha + \beta + \gamma - \pi .
\end{equation}
Summing the solid-angles corresponding to all the 12 triangles in the
cubes surface, we obtain $\sum \Theta = 4\pi\cdot n$ where $n$ is
an integer taken to be the charge of the monopole inside the lattice cube. 
$|n|$  has an upper bound of $5$ but in practice we rarely observe charges larger
that $2$ (which we interpret as two coincident unit charges).

\end{appendix}

\end{document}